\documentstyle[12pt,a41,axodraw]{article}

\newcommand\beq{\begin{equation}}
\newcommand\eeq{\end{equation}}
\newcommand\bea{\begin{eqnarray}}
\newcommand\eea{\end{eqnarray}}

\begin{document}
\thispagestyle{empty}

\vspace*{\fill}
\begin{center}
{\LARGE\bf Higher order QCD Corrections to} 
\vspace{2mm}

{\LARGE\bf Fragmentation Functions}

\vspace{2mm}
{\LARGE\bf  in $e^+~e^-$ Annihilation}

\vspace*{20mm}
\large
{W.L. van Neerven \footnote{Talk presented at the Zeuthen Workshop
on Elementary Particle Theory "Loops and Legs in Gauge Theories"
Rheinsberg, Germany, April 19-24, 1998.}}
\\

\vspace{2em}

\normalsize
{\it DESY-Zeuthen, Platanenallee 6, D-15738 Zeuthen, Germany}%
\footnote{
On leave of absence from Instituut-Lorentz, University of Leiden,
P.O. Box 9506, 2300 RA Leiden, The Netherlands.}

\vspace{4cm}
\end{center}
\vspace*{\fill}
\begin{abstract}
\noindent
We analyze the second order QCD corrections to the fragmentation
functions
$F_k^{\rm H}(x,Q^2)$ ($k=T,L,A$) which are measured in $e^+~e^-$
annihilation
From these fragmentation functions one can derive the integrated
transverse ($\sigma_T$), longitudinal ($\sigma_L$) and asymmetric
($\sigma_A$)
cross sections. The sum $\sigma_{tot}=\sigma_T+\sigma_L$ corrected up
to order $\alpha_s^2$ agrees with the well known result in the
literature.
It turns out that the order $\alpha_s^2$ corrections to the transverse
and asymmetric quantities are small in contrast to our findings for
$F_L^{\rm H}(x,Q^2)$ and $\sigma_L$ where they turn out to be large.
Therefore in the latter case one gets a better agreement between the
theoretical predictions and the data obtained from he LEP experiments.
\end{abstract}
\vspace*{1cm}

\section{Introduction}
\begin{figure}
\begin{center}
  \begin{picture}(185,130)(0,0)
    \ArrowLine(130,80)(180,120)
    \DashArrowLine(130,70)(185,45){5}
    \DashArrowLine(130,62)(195,20){5}
    \DashArrowLine(130,55)(175,15){5}
    \Line(180,65)(195,20)
    \Line(155,10)(195,20)
    \GCirc(120,70){20}{0.3}
    \Photon(40,70)(100,70){3}{7}
    \ArrowLine(0,40)(40,70)
    \ArrowLine(0,100)(40,70)
    \Text(0,110)[t]{$e^-$}
    \Text(0,38)[t]{$e^+$}
    \Text(190,130)[t]{$H$}
    \Text(80,90)[t]{$V$}
    \Text(208,25)[t]{$'{\rm X}'$}
    \Text(155,115)[t]{$p,s$}
    \Text(25,102)[t]{$l_1,\sigma_1$}
    \Text(25,48)[t]{$l_2,\sigma_2$}
    \Text(80,65)[t]{$\rightarrow q$}
  \end{picture}
  \caption[]{Kinematics of electron positron annihilation
            $e^- + e^+ \rightarrow H  + '{\rm X}'$ }
  \label{fig:1}
\end{center}
\end{figure}
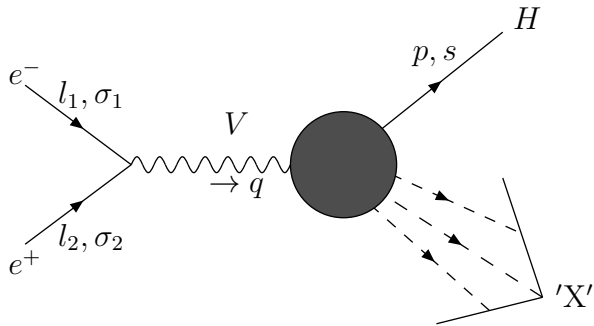

Semi inclusive hadron production in $e^+~e^-$ annihilation into a vector 
boson V ($V=\gamma,Z$) proceeds via the reaction (see \ref{fig:1})
\begin{eqnarray}
\label{eq1}
  e^-(l_1,\sigma_1) + e^+(l_2,\sigma_2) \rightarrow 
V(q) \rightarrow H(p,s) + ``X" \,.
\end{eqnarray}
In the process above $``X"$ denotes any inclusive final hadronic state and 
$H$ represents
either a specific charged outgoing hadron or a sum over all charged hadron
species. 
The unpolarized differential cross section of the above process is given by
\begin{eqnarray}
\label{eq2}
  \frac{d^2\sigma^H(x,Q^2)}{dx\,d\cos\theta} &=& \frac{3}{8}
(1+\cos^2\theta)\frac{d\sigma_T^H(x,Q^2)}{dx}
  + \frac{3}{4}\sin^2\theta\frac{d\sigma_L^H(x,Q^2)}{dx} 
\nonumber\\[2ex]
&& + \frac{3}{4}\cos\theta\frac{d\sigma_A^H(x,Q^2)}{dx}.
\end{eqnarray}
The transverse, longitudinal
and asymmetric cross sections are given by $\sigma_T^H$,
$\sigma_L^H$, and $\sigma_A^H$ respectively. The latter, which is due to
parity violation, only shows up if the intermediate vector boson is given
by the $Z$-boson and is absent when $V=\gamma$.
The cross sections depend in addition to the
CM energy $Q$ also on the Bj{\o}rken 
scaling variable $x$ defined by
\begin{eqnarray}
\label{eq3}
  x = \frac{2pq}{Q^2},\hspace{8mm} q^2 = Q^2 > 0,\hspace{8mm} 0 < x \leq 1\,.
\end{eqnarray}
In the centre of mass (CM) frame of the
electron-positron
pair this variable can be interpreted as the fraction of the beam energy
carried away
by the hadron $H$. The variable $\theta$ denotes the angle of emission of
particle $H$
with respect to the electron beam direction in the CM frame.
Before the advent of LEP1 the CM energies were so low ($Q \ll M_Z$) that
$\sigma_A$ could not be measured and no effort was made to separate $\sigma_T$
{}from $\sigma_L$ so that only data for $\sigma_T+\sigma_L$ were available.
After LEP1 came into operation one was able to measure
$\sigma_T$ and $\sigma_L$ separately \cite{busk}, \cite{akers}. Moreover
$\sigma_A$ could be determined for the first time \cite{akers}, \cite{abreu}.
The separation of $\sigma_T$ and $\sigma_L$
is important because the latter cross section enables us to extract the strong
coupling constant $\alpha_s$ and allows us to determine the gluon fragmentation
density $D_g^H(x,\mu^2))$ with a much higher accuracy as could be done before.
Furthermore the measurement of $\sigma_A$ provides us with information
on hadronization effects since the QCD corrections are very small.
In the QCD improved parton model, which describes the production of the parton
and its subsequent fragmentation into a hadron $H$, the cross sections
$\sigma_k^H$ ($k=T,L,A$)
can be expressed as follows
\begin{eqnarray}
\label{eq4}
&& \frac{d\sigma_k^H(x,Q^2)}{dx} = \int_x^1\,\frac{dz}{z}\Big [
\sigma_{\rm tot}^{(0)}(Q^2)\,\Big \{ D_S^H (\frac{x}{z},\mu^2)
    {\cal C}_{k,q}^S(z,Q^2/\mu^2)
  + D_g^H (\frac{x}{z},\mu^2) \cdot
\nonumber\\[2ex]
&& \cdot {\cal C}_{k,g}^S (z,Q^2/\mu^2)\Big \}
  +\sum_{f=1}^{n_f}\,\sigma_f^{(0)}(Q^2)\,
  D_{NS,f}^H\left(\frac{x}{z},\mu^2\right)
  {\cal C}_{k,q}^{NS}(z,Q^2/\mu^2)\Big ],
\end{eqnarray}
for $k=T,L$. In the case of the asymmetric cross section we have
\begin{eqnarray}
\label{eq5}
\frac{d\sigma_A^H(x,Q^2)}{dx} = \int_x^1\,\frac{dz}{z}\Big [\sum_{f=1}^{n_f}\,
A_f^{(0)}(Q^2)
D_{A,f}^H(\frac{x}{z},\mu^2)
{\cal C}_{A,q}^{NS}(z,Q^2/\mu^2)\Big ].
\end{eqnarray}
In the formulae above, which only hold for massless quarks, we have introduced 
the following notations.
The functions $D_l^H(z,\mu^2)$ ($l=q,\bar q,g$), which depend on the 
factorization/renormalization scale $\mu$, stand for the parton fragmentation 
densities. The singlet (S) and  non-singlet (NS,A) combinations with respect
to the flavour group $SU(n_f)$ are defined by
\begin{eqnarray}
\label{eq6}
D_S^H(z,\mu^2) & = & \frac{1}{n_f}\,\sum_{q=1}^{n_f}\Big (D_q^H(z,\mu^2) +
    D_{\bar{q}}^H(z,\mu^2)\Big ) \,,
\nonumber\\[2ex]
D_{NS,q}^H(z,\mu^2) &=& D_q^H(z,\mu^2) + D_{\bar{q}}^H(z,\mu^2) 
- D_S^H(z,\mu^2) \,,
\nonumber\\[2ex]
D_{A,q}^H(z,\mu^2)& =& D_q^H(z,\mu^2) - D_{\bar{q}}^H(z,\mu^2) \,.
\end{eqnarray}
The index $q$ stands for the quark species and $n_f$ denotes the number of 
light flavours.
The pointlike cross
section $\sigma_q^{(0)}$ and the asymmetry factor $A_q^{(0)}$ of the process 
$e^+ + e^- \rightarrow q + \bar{q}$ can be found in \cite{alt}, \cite{rn}.
The total cross section, summed over all flavours is
given by $\sigma_{\rm tot}^{(0)}(Q^2) = \sum_{q=1}^{n_f}\sigma_q^{(0)}(Q^2)$.
The QCD corrections in eqs. (\ref{eq4}), (\ref{eq5}) are described by the 
coefficient functions ${\cal C}_{k,l}^r$
($k=T,L,A$; $l=q,g$). Like the fragmentation densities they depend on the scale
$\mu$ and they can be split into a singlet ($r=S$) and a non-singlet 
part ($r=NS$). 
From (\ref{eq2}) we can derive the total hadronic cross section
\begin{eqnarray}
\label{eq7}
\sigma_{\rm tot}(Q^2) &=& 
\frac{1}{2}\sum_H\,\int_0^1dx\,\int_{-1}^1 d\cos\theta\,
\Big (x\frac{d^2\sigma^{\rm H}(x,Q^2)}{dx\,d\cos\theta}\Big ) 
\nonumber\\[2ex]
&=& \sigma_T(Q^2) + \sigma_L(Q^2) \,,
\end{eqnarray}
with
\begin{eqnarray}
\label{eq8}
\sigma_k(Q^2) = \frac{1}{2}\sum_H\,\int_0^1dx\,x\,\frac{d\sigma_k^{\rm H}
(x,Q^2)}{dx}\,,
\hspace*{7mm} k=T,L \,,
\end{eqnarray}
where one has summed over all types of outgoing hadrons $H$.
From the momentum conservation sum rule given by
\begin{eqnarray}
\label{eq9}
\sum_H\,\int_0^1 dx\, x\, D_l^{\rm H}(x,\mu^2) = 1 \,, 
\qquad  l=q,\bar q,g \,,
\end{eqnarray}
and Eqs. (\ref{eq4}), (\ref{eq8}) one can derive
\begin{eqnarray} 
\label{eq10}
\sigma_k(Q^2) = \sigma_{\rm tot}^{(0)}(Q^2)\int_0^1dx\,x\Big [
{\cal C}_{k,q}^{\rm S}(x,Q^2/\mu^2)
+ \frac{1}{2}{\cal C}_{k,g}^{\rm S}(x,Q^2/\mu^2)\Big ]\,.
\end{eqnarray}
Finally we also define the transverse, longitudinal and asymmetric 
fragmentation functions
$F_k^H(x,Q^2)$
\begin{eqnarray}
\label{eq11}
F_k^{\rm H}(x,Q^2) = \frac{1}{\sigma_{\rm tot}^{(0)}(Q^2)}\,
\frac{d\sigma_k^{\rm H}(x,Q^2)}{dx} \,,
\hspace{4mm} k=(T,L,A) \,.
\end{eqnarray}
One observes that
the above fragmentation functions
\footnote{Notice that we make a distinction in nomenclature
between
the fragmentation densities $D_q^H$,$D_g^H$ and the fragmentation functions
$F_k^H$.}
are just the timelike analogues of the structure functions measured in deep 
inelastic electron-proton scattering.
\section{Order $\alpha_s^2$ corrected coefficient functions}
The coefficient functions ${\cal C}_{k,l}^r$ corrected up to order $\alpha_s^2$
receive contributions from the following parton subprocesses. 
In zeroth order we have the Born reaction
\begin{eqnarray}
\label{eq12}
V \rightarrow "q" + \bar q \,,
\end{eqnarray} 
where $"l"$ ($l=q,\bar q,g$) denotes the detected parton which subsequently
fragments into the hadron of species $H$.
In next-to-leading order (NLO) one has to compute the one-loop virtual  
corrections to reaction (\ref{eq12}) and the parton subprocesses
\begin{eqnarray}
\label{eq13}
V \rightarrow  "q" + \bar q + g \,,
\end{eqnarray}
\begin{eqnarray}
\label{eq14}
V \rightarrow  "g" + q + \bar q \,.
\end{eqnarray}
After mass factorization of the collinear divergences which arise in the above
processes one obtains the coefficient functions which are presented in
\cite{alt}.
The determination of the order $\alpha_s^2$ contributions involves the 
computation of the two-loop corrections to (\ref{eq12}) and the one-loop  
corrections to eqs. (\ref{eq13}),(\ref{eq14}). Furthermore one has to calculate
 the following subprocesses
\begin{eqnarray}
\label{eq15}
V \rightarrow  "q" + \bar q + g + g \,,
\end{eqnarray}
\begin{eqnarray}
\label{eq16}
V \rightarrow  "g" + q + \bar q + g \,,
\end{eqnarray}
\begin{eqnarray}
\label{eq17}
V \rightarrow  "q" + \bar q + q + \bar q  \,.
\end{eqnarray}
In reaction (\ref{eq17})) the two anti-quarks, which are inclusive, 
can be identical as
well as non-identical. Notice that in the above reactions the detected quark 
can be replaced by the detected anti-quark so that in reaction (\ref{eq17}) one 
can also distinguish between the final states containing identical quarks and
non-identical quarks. After mass factorization and renormalization for which 
we have chosen the ${\overline {\rm MS}}$-scheme one obtains the order
$\alpha_s^2$ contributions to the coefficient functions which are presented in
\cite{rn}.
\section{Review of the most important results}
The most important results of our calculations can be summarized as follows.
From eq. (\ref{eq10}) and the coefficient functions originating from the 
processes above
we can obtain $\sigma_L$ and $\sigma_T$ corrected up to order $\alpha_s^2$
\begin{eqnarray}
\label{eq18}
&& \sigma_T(Q^2) =  \sigma^{(0)}_{\rm tot}(Q^2)\Big [ 1 + 
\Big (\frac{\alpha_s(\mu^2)}{4\pi}\Big )^2 \Big ( C_F^2\Big \{6\Big \}
\nonumber\\[2ex]
&& + C_AC_F\Big \{
    -\frac{196}{5}\zeta(3) - \frac{178}{30}\Big \} 
  + n_fC_FT_f\Big \{ 16\zeta(3) + \frac{8}{3}\Big \}\Big )\Big ]\,,
\end{eqnarray}
\begin{eqnarray}
\label{eq19}
&& \sigma_L(Q^2) =  \sigma^{(0)}_{\rm tot}(Q^2)\Big [\frac{\alpha_s(\mu^2)}{
4\pi}
C_F\Big \{3\Big \} + \Big (\frac{\alpha_s(\mu^2)}{4\pi}\Big )^2 \Big (
    C_F^2\Big \{-\frac{15}{2}\Big \} 
\nonumber\\[2ex]
&& + C_AC_F\Big \{
   -11 \ln\frac{Q^2}{\mu^2}
  -\frac{24}{5}\zeta(3) + \frac{2023}{30}\Big \} + n_fC_FT_f\Big \{
  4\ln\frac{Q^2}{\mu^2} - \frac{74}{3}\Big \}\Big )\Big ]\,.
\nonumber\\
\end{eqnarray}
Addition of $\sigma_L$ and $\sigma_T$ yields the well known answer for
$\sigma_{\rm tot}$ (\ref{eq7}) which is in agreement with the literature 
\cite{ckt}.
Hence this quantity provides us with a check on our calculation of the
longitudinal and transverse coefficient functions. Notice that in lowest order
$\sigma_{\rm tot}$ only receives a contribution from the transverse cross
section whereas the order $\alpha_s$ contribution can be only attributed
to the longitudinal part. In order $\alpha_s^2$ both $\sigma_L$
and $\sigma_T$ contribute to $\sigma_{\rm tot}$.\\
Because of the high sensitivity of expression (\ref{eq19}) to the value 
of $\alpha_s$,
the longitudinal cross section provides us with an excellent tool to measure the
running coupling constant.
To illustrate the importance of the order $\alpha_s^2$ contribution to
$\sigma_L$ we have computed the ratio 
\begin{eqnarray}
\label{eq20}
R_L(Q^2)=\frac{\sigma_L(Q^2)}{\sigma_{tot}(Q^2)} \,,
\end{eqnarray}
for $\mu=Q=M_Z$ and $\alpha_s(5,M_Z)=0.126$. The result is
\begin{eqnarray}
\label{eq21}
R_L = 0.040 + 0.014 = 0.054\,, \qquad (0.057 \pm 0.005) \,,
\end{eqnarray}
where the first and the second number represent the order $\alpha_s$ and
order $\alpha_s^2$ contribution respectively. Between the brackets we have 
quoted the result from OPAL \cite{akers}. 
Here one observes a considerable improvement when the order
$\alpha_s^2$ contributions are included. Recently DELPHI \cite{abreu} used
Eq. (\ref{eq19}) to determine the strong coupling constant. Their 
measurement of $R_L$ yields
\begin{eqnarray}
\label{eq22}
R_L = 0.051 \pm 0.01\, (stat.) \pm 0.007\, (syst.) \,,
\end{eqnarray}
from which the strong coupling constant can be extracted. The result is
\begin{eqnarray}
\label{eq23}
\alpha_s^{\rm NLO}(5,M_Z) = 0.120 \pm 0.002\, (stat.) \pm 0.013\, (syst.) \,.
\end{eqnarray}
If one includes power corrections to $\sigma_L$ which are due to higher
twist contributions of the order $\Lambda/Q$ (see \cite{bbm}) then
one obtains
\begin{eqnarray}
\label{eq24}
\alpha_s^{\rm NLO+POW}(5,M_Z) = 0.101 \pm 0.002\, (stat.) \pm 0.013\, (syst.)
\pm 0.007\, (scale) \,,
\nonumber\\
\end{eqnarray}
where the scale uncertainty comes from varying the renormalization
scale in he range $0.5~Q < \mu < 2~Q$.
The result above shows that the power corrections enhance $\sigma_L$ so
that the strong coupling constant decreases. A similar analysis (see 
\cite{rn}) shows that also the longitudinal fragmentation function
$F_L$ receives large order $\alpha_s^2$ contributions. However the 
corrections to the transverse and asymmetric fragmentation functions
are small.

Finally we discuss the order $\alpha_s^2$ corrections to the asymmetric
cross section. The latter is given by
\begin{eqnarray}
\label{eq25}
\sigma_A(Q^2) = \sum_H\,\int_0^1\,dx\,\frac{d\sigma_A^{\rm H}
(x,Q^2)}{dx}= Q_A \int_0^1 \,dx\, {\cal C}_{A,q}^{\rm NS}(x,Q^2/\mu^2)\,,
\end{eqnarray}
Up to NNLO he above expression yields the following result
\begin{eqnarray}
\label{eq26}
\sigma_A=Q_A \Big [ 1 + \Big (\frac{\alpha_s(\mu^2)}{4\pi}\Big )^2 \Big \{
-12 \beta_0 \zeta(3) \Big \} \Big ]\,, \qquad \beta_0=\frac{11}{3}C_A -
\frac{4}{3}T_f n_f \,.
\nonumber\\
\end{eqnarray}
Notice that $\sigma_A$ is independent of the factorization scale. However
it depends like $\sigma_L$ and $\sigma_A$ on the renormalization scale
via $\alpha_s$. From Eq. (\ref{eq26}) we infer that the NLO correction is
zero but the NNLO contribution is nonvanishing although numerically it
is very small for $\mu=Q=M_Z$. 
Finally in \cite{rn} we have evaluated the first and 
second moment of the asymmetric structure function and compared it
with the data. The NNLO result for the first moment is
\begin{eqnarray}
\label{eq27}
\int_{0.1}^1 dx F_A(x,M_Z^2) = -0.016\,, \quad &-0.0229 \pm 0.0044&
\, (OPAL) \,,
\nonumber\\[2ex] 
& -0.028 \pm 0.006& \, (DELPHI) \,.
\nonumber\\
\end{eqnarray}
The second moment becomes
\begin{eqnarray}
\label{eq28}
\int_{0.1}^1 dx \frac{x}{2} F_A(x,M_Z^2) = -0.0020 \,,
\quad &-0.00369 \pm 0.00046&\, (OPAL) \,,
\nonumber\\[2ex]
& -0.0036 \pm 0.0008&\, (DELPHI) \,.
\nonumber\\
\end{eqnarray}
We also computed the above sum rules up to NLO. However
there is hardly any difference between NLO and NNLO which could already 
be expected from the comments made below Eq. (\ref{eq26}).
The numbers presented above reveal that the experimental values are far
below the theoretical predictions. Furthermore it turns out that the LO
results are much closer to the ones obtained from experiment. In the
case of Eq. (\ref{eq27}) we obtain $-0.023$ whereas for Eq. (\ref{eq28})
we get $-0.0027$. Probably higher twist corrections might become very
important as we saw in the case of $\sigma_L$ below (\ref{eq23}).

\end{document}